\documentclass{aa}
\usepackage{epsfig,amsmath,amssymb}

\def\mjup{M_{\rm J}}
\def\rjup{R_{\rm J}}
\def\mearth{\,{\rm M}_\oplus}

\def\lbol{L_{\rm bol}}
\def\simgr{\,\hbox{\hbox{$ > $}\kern -0.8em \lower 1.0ex\hbox{$\sim$}}\,}
\def\simle{\,\hbox{\hbox{$ < $}\kern -0.8em \lower 1.0ex\hbox{$\sim$}}\,}
\def\beq{\begin{equation}}
\def\eeq{\end{equation}}

\begin{document}


\title{Hot-Jupiters and hot-Neptunes: a common origin?
}
 \author{I. Baraffe\inst{1,2}, G. Chabrier\inst{1}, T. S. Barman\inst{3}, F. Selsis\inst{1}, F. Allard\inst{1} and P.H. Hauschildt\inst{4} 
}

\offprints{I. Baraffe}

\institute{C.R.A.L (UMR 5574 CNRS),
 Ecole Normale Sup\'erieure, 69364 Lyon
Cedex 07, France (ibaraffe, chabrier, fselsis, fallard@ens-lyon.fr)
\and
International Space Science Institute, Hallerstr. 6, CH-3012, Bern, Switzerland
\and
Department of Physics and Astronomy, UCLA, Los Angeles, CA 90095, USA
(barman@astro.ucla.edu)
\and
Hamburger Sternwarte, Gojenbergsweg 112,
21029 Hamburg, Germany (yeti@hs.uni-hamburg.de)
}

\date{Received /Accepted}

\titlerunning{Hot-Jupiters and hot-Neptunes: a common origin?}
\authorrunning{Baraffe et al.}
\abstract{We compare evolutionary models for close-in exoplanets coupling irradiation
and evaporation due respectively to the thermal and high energy flux of the parent star with 
observations of recently discovered new transiting planets. The models
provide an overall good agreement with observations, although at the very limit of the quoted
error bars of OGLE-TR-10, depending on its age.
Using the same general theory,
we show that the three recently detected hot-Neptune planets (GJ436, $\rho$ Cancri, $\mu$ Ara)
may originate from 
more massive gas giants which have undergone significant evaporation. We thus suggest
that hot-Neptunes and hot-Jupiters may share the same origin and evolution history.
Our scenario provides testable predictions in terms of the mass-radius relationships
of these hot-Neptunes.
}

\maketitle

\section{Introduction}
Determining the evolution and formation mechanism of extrasolar planets in very
small orbits, one to two hundred times smaller than Jupiter's orbit, remains a
challenging puzzle.  This puzzle recently became even more complex when radial
velocity surveys of unprecedented accuracy discovered three extrasolar planets
similar in mass to Neptune, i.e. 14 to 21 $\mearth$, with orbital periods
$P$=2.6-9.5 days and separations $a$=0.028-0.09 AU (Santos et al. 2004, McArthur et al. 2004, Butler et al. 2004).
An emerging short-period,
Neptune-mass, exoplanet population presents us with intriguing questions about
their compositions and origins, with potential implications for the
understanding of our own solar system. Because of the presence of  Jupiter-mass planets  at orbital separations even closer than that of these hot-Neptunes,
people ruled out the possibility that the latter were born as more massive giant planets which have lost a significant fraction of their envelope. 
The current belief following this intriguing discovery
 was then that these hot-Neptunes are of different nature 
and have different formation and evolutionary histories than their more
massive counterparts (Santos et al. 2004, McArthur et al. 2004, Mazeh et al. 2004).

In a recent paper, we explored the effects of evaporation on the
evolution of  short-period jovian planets, using a consistent treatment of
the interior structure and the irradiated atmospheric structure
(Barman et al. 2001, Chabrier et al. 2004, Baraffe et al. 2004).  
The idea of evaporation is supported by the recent
discovery of an extended atmosphere around the transiting exoplanet HD 209458b
(Vidal-Madjar et al. 2003, Vidal-Madjar et al. 2004), demonstrating the occurrence of strong atmospheric evaporation and mass loss for short-period irradiated planets. 
These models could successfully reproduce the observed radius
of the transit OGLE-TR-56 (Chabrier et al. 2004) but the case of HD 209458b
remains difficult to explain (Baraffe et al. 2003), raising some contention over the general
applicability of current evolution theories to hot-Jupiters.

In this Letter, we first show that our models including irradiation effects from the thermal and the high energy fluxes of the parent star are consistent with  recent observations of four other new transits, confirming the peculiarity of HD 209458b.
Motivated by this success and by the
recent discovery of Neptunian exoplanets, we have 
extended our calculations to the evolution of
irradiated and evaporating gaseous planets
below the mass of Saturn, down to masses comparable to those of Uranus and
Neptune. 
We show that,
contrary to present belief,  all of the recent neptunian objects, along
with their short-period Jupiter-mass cousins, could have originated from more
massive irradiated gas giants which suffered from evaporation induced by the
parent star's high-energy radiation.  We also present
characteristic mass-radius relationships that bear the imprint of irradiation
and atmospheric evaporation on a planet's evolution.  These predictions will be
tested in the very near future by surveys capable of detecting Earth-sized
transiting planets.

\section{Analysis of newly detected transit planets}

\begin{figure}
\psfig{file=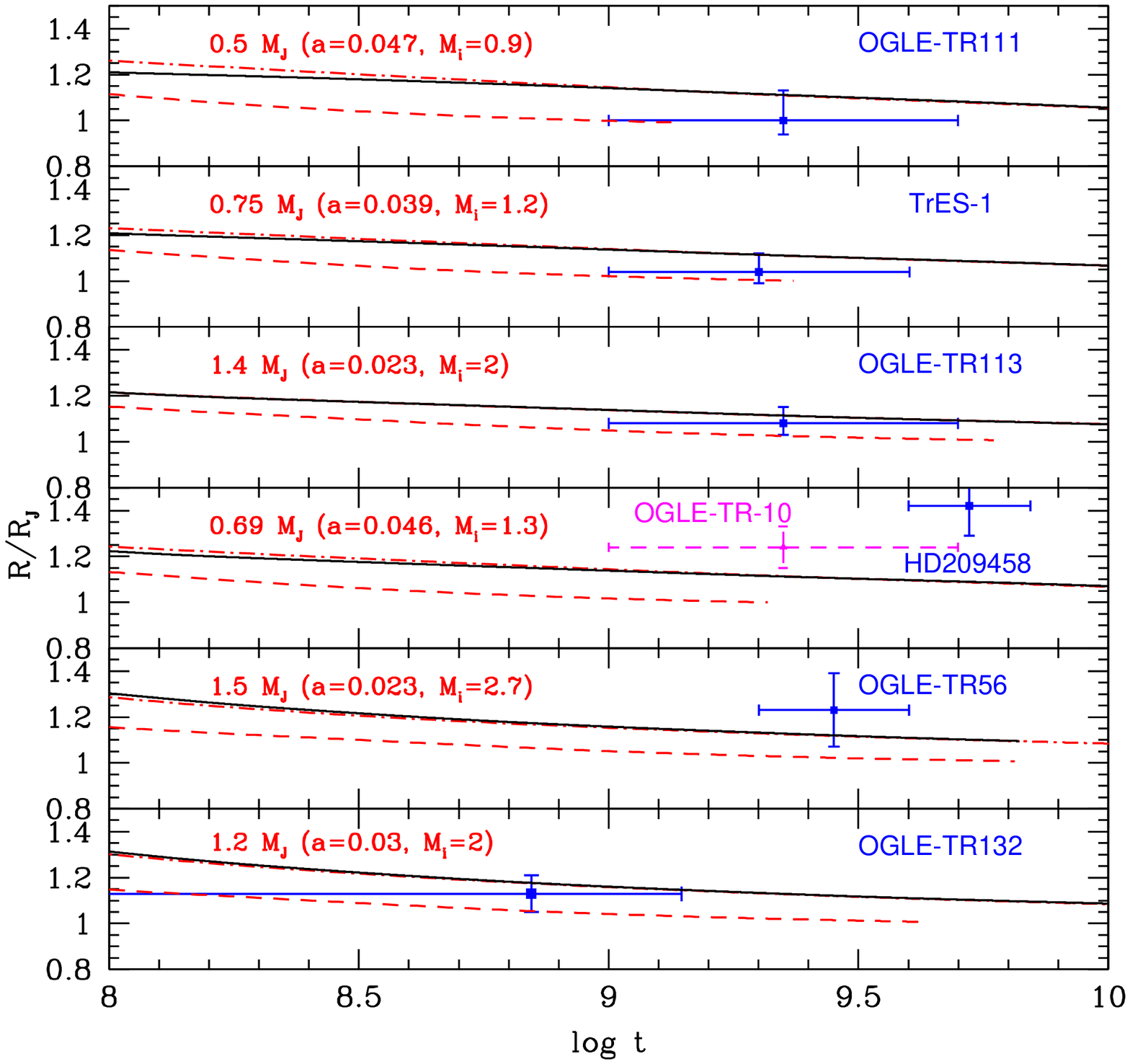,height=120mm,width=88mm} 
\caption{Evolution of the radius (in $\rjup$) for currently known transiting
planets. The incident stellar flux contribution, $F_{\rm inc}$, increases from top to bottom.  The measured
mass and orbital separation are indicated in each panel. The curves correspond
to evolutionary models for three cases: without irradiation (dashed lines),
with irradiation effects on the thermal structure but no evaporation
(dash-dotted lines), and with irradiation and evaporation effects (solid
lines). For the latter cases, the initial mass $M_i$ is also indicated
in each panel. 
 Note that in most cases the dash-dotted (no evaporation)
 and solid (evaporation) lines are undistinguishable (see text).}
\label{fig1}
\end{figure}

Since the discovery of the two first and well confirmed transits HD 209458b (Charbonneau et al. 2000)  and OGLE-TR-56b (Konacki et al. 2003), five new candidates were found very recently, providing more stringent
constraints on the theory of irradiated planets. 
The properties of these new transits and of their parent star are
summarized in Table 1, which also provides the incident flux $F_{\rm inc}$ irradiating the planet, as defined in Baraffe et al. (2003)\footnote{ 
$F_{\rm inc} = {1 \over 2} \, ({R_\star \over a})^2 F_\star$, 
where ${R_\star}$ and $F_\star$ are respectively the radius and  the total flux of
the parent star, and $a$ the orbital separation.}.
Coupling consistently irradiated atmosphere models calculated with
the appropriate incident fluxes (Barman et al. 2005) and interior structures,  we have calculated
the evolution of planets
with parameters  ($a$, $M_{\rm_p}$, $F_{\rm inc}$) characteristic of each of the new observed transits. The evolutionary models and their input physics are described in Baraffe et al. (2003) and references therein. Figure \ref{fig1}  shows a comparison between evolutionary models and observations
for all transiting planets. OGLE-TR-10b and HD 209458b are displayed in the same panel, since they have similar properties, in terms of planet mass and incident flux,  and the same model is appropriate to describe their evolution. When the age of the system is undetermined, we adopt an arbitrary
age ranging from 1 to 5 Gyr.

Besides irradiation, we also take into account the effect of evaporation of the planet gaseous content
due to the stellar incident high energy flux, based on
the hydrodynamic model of atmospheric evaporation developed recently 
by Lammer et al. (2003) and applied by 
Baraffe et al. (2004) to the case of the two first transits HD 209458b and OGLE-TR-56b.
The description of the evaporation model and the details of its coupling with the planet evolution
can be found in the two aforementioned references. Note that the evaporation rate is based
 on heating by the {\it age-dependent}
stellar XUV and Lyman-$\alpha$ radiation, calibrated to {\it observed} age-luminosity
$L_{XUV}(t)/\lbol(t)$ relationships (Lammer et al. 2003, Ribas et al. 2005), and takes root on previous studies devoted
to escaping atmospheres of the Solar system terrestrial planets 
(\"Opik 1963, Watson et al. 1981).
Such hydrodynamic evaporation yields mass
loss rates in agreement with the observationally determined {\it lower}
limit for HD 209458b: $\dot M\gtrsim 10^{10}$ g s$^{-1}\approx
10^{-13}\,\mjup$ yr$^{-1}$, at the present epoch (Vidal-Madjar et al. 2003).

Models including irradiation and evaporation are shown by the solid lines in Fig. \ref{fig1},
with initial masses of the
planets ranging between 0.9 and 2.7 $\mjup$.  These initial conditions remain uncertain,
because of the uncertainties in the evaporation model, 
but the results hold, qualitatively, with different evaporation parameters.
Figure \ref{fig1} shows that consistent irradiated models, including irradiation effects due to both thermal and high energy fluxes of the parent star, are in
excellent agreement with the observed radii at the proper ages for all of the
planets except HD 209458b and possibly OGLE-TR-10b, depending on its age. 
Figure \ref{fig1} also shows that models with evaporation (solid lines) are almost undistinguishable
from the case without evaporation (dash-dotted lines). This
 illustrates the concept of critical mass defined in Baraffe et al. (2004), below which a 
gaseous planet  should evaporate entirely after a certain age, for a given incident flux. 
The evolution of a planet above this
critical mass, however, remains
unaffected by evaporation up to several billion years.  All the progenitors of the
transiting planets fall in this latter domain except, interestingly enough,
HD 209458b and OGLE-TR-10b, which are just below this limit. 
The specific case of HD 209458b was already discussed in Baraffe et al. (2003) and Chabrier et al. (2004). These authors demonstrated, among possible explanations for the particularly large radius, that a mechanism, such e.g. as the one suggested by Guillot and Showman (2002), which would
 dissipate a fraction as small as 0.1\%-0.5\% of the stellar incident flux at the planet internal adiabat level could reproduce
the observations of both HD 209458b and OGLE-TR-56b. We have presently tested
this hypothesis on OGLE-TR-111, which is the least massive and the {\it least irradiated} planet (see Table 1).
Adding 0.1\% of the incident stellar energy to the corresponding irradiated model yields  a radius larger by 7\% than the upper limit of the observed radius. If a dissipative mechanism, still to be identified, is
a common feature of all close-in planets, our test indicates that either it has to be rather inefficient
in OGLE-TR-111 or this planet has a significant rocky or icy core, yielding a $\sim 5\%$ smaller radius than a
purely gaseous object (Saumon et al. 1996).

\begin{table*}
\caption{ Properties of transiting extrasolar planets and of their parent star. The following quantities are displayed: the orbital separation, the mass and radius of the planet, the mass, effective temperature,
spectral type and radius of the star, the incident flux (see text) and the age when determined.}
\label{tab1}
\begin{tabular}{lccccccccc}
\hline
\hline
 Object & a(au) & M$_p$(M$_{Jup}$) & R$_p$(R$_{Jup}$) &  
 M$_{\star}$ (M$_{\odot}$) & T$_{eff \star}$ (K) & Sp. type & R$_{\star}$(R$_{\odot}$) & $\log(F_{\rm inc})$ & age (Gyr) \\
 \hline
 \hline
  & & &  & & & &  &    & \\
\vspace{1eX}
  OGLE-TR-132$^a$  & 0.0306 & 1.19$\pm 0.13$ & 1.13 $\pm 0.08$    & 1.35 & 6411 &  F & 1.43  & 9.35  & 0-1.4 \\
\vspace{1eX}
  OGLE-TR-56$^b$   & 0.023 & 1.45$\pm 0.23$ & 1.23 $\pm 0.16$    & 1.04 & 6000  &G & 1.10  & 9.26 & 3 $\pm 1$ \\
\vspace{1eX}
  HD 209458$^c$    & 0.046  & 0.69$\pm 0.02$  & 1.42$^{+0.10}_{-0.13}$    & 1.06 & 6000  &G & 1.18 & 8.72  & 4-7 \\
\vspace{1eX}
  OGLE-TR-10$^d$    & 0.042  & 0.57$\pm 0.12$  & 1.24$\pm 0.09$   & 1 & 5800  &G & 1 & 8.60 &  - \\
\vspace{1eX}
  OGLE-TR-113$^e$  & 0.023 & 1.35$\pm 0.22$ & 1.08$^{+0.07}_{-0.05}$    & 0.77 & 4750 &K & 0.765 & 8.53 & - \\
\vspace{1eX}
  TrES-1$^f$      & 0.039 & 0.76$\pm 0.05$ & 1.04$^{+0.08}_{-0.05}$    & 0.89 & 5250 &K & 0.83 & 8.32& 2.5 $\pm$ 1.5\\
\vspace{1eX}
  OGLE-TR-111$^g$  & 0.047  & 0.53$\pm 0.11$ & 1.00$^{+0.13}_{0.06}$    & 0.82 & 5070 &G-K & 0.85  & 8.12  & - \\
 \hline
 \end{tabular}
\\
$^a$Moutou et al. 2005 \hskip 0.15cm
$^b$Torres et al. 2004 \hskip 0.15cm
$^c$Cody \& Sasselov 2002 \hskip 0.15cm
$^d$Konacki et al. 2005 \hskip 0.15cm
$^e$Bouchy et al. 2004 \\
$^f$Sozzetti et al. 2004 \hskip 0.15cm
$^g$Pont et al. 2004

\end{table*}

\section{The case of hot-Neptunes}

Current transits do not provide constraints on the evaporation rates, 
since, in their mass range, their mass-radius relationship remains unaffected by the process of evaporation. However, the good agreement of present models
with most of the currently observed transits gives us confidence on a scenario
based on a {\it global} process of irradiation and evaporation. We have thus extended this 
scenario to initial masses below the critical mass in order to explore the regime
of sub-jovian mass planets. 
 
\begin{figure}
\psfig{file=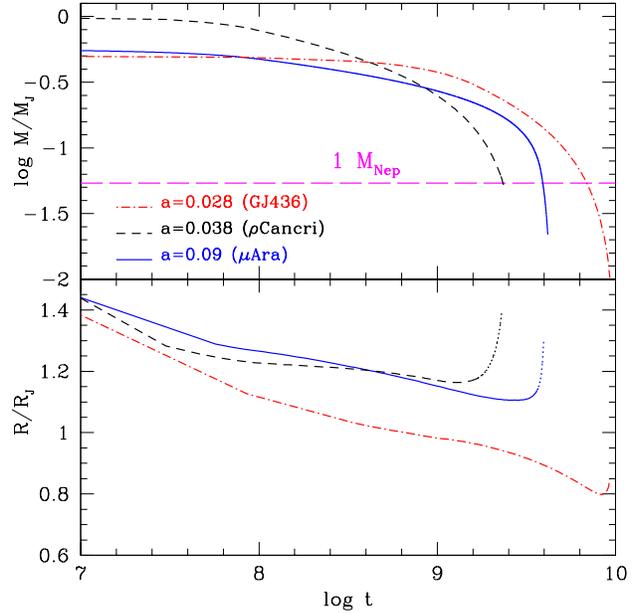,height=88mm,width=88mm} 
\caption{Evolution of evaporating giant planets reaching a Neptune-mass (indicated in
the top panel by the long-dashed line, in Jupiter-mass units $\mjup$) within a
few Gyr.  Three planet-star cases are shown for: 
the G5 star $\mu$ Ara  (Santos et al. 2004) (solid),
the G8 star $\rho^1$ Cancri (McArthur et al. 2004) (dashed), and
the M2.5 star GJ 436  (Butler et al. 2004) (dash-dotted).
The planet's orbital separation $a$ (in AU) for each system is indicated in the
upper panel.}
\label{fig2}
\end{figure}

Figure \ref{fig2} displays the evolution of hydrogen/helium planets with initial masses
ranging from 0.5 to 1 $\mjup$ orbiting a parent star under conditions characteristic
of the three newly discovered Neptunian planets, namely 0.028 AU from an M2.5
star (GJ436, Butler et al. 2004), 0.038 AU from a G8 star ($\rho^1$Cancri,
McArthur et al. 2004) and 0.09 AU from a G5 star ($\mu$Ara, Santos et al. 2004).
The observed spectra and magnitudes of the parent stars, complemented by
well established age-activity relations for similar type stars,
imply that these
systems must be a few billion year old. After $\sim 2$-5 Gyr, these planets
have lost more than 90\% of their initial mass, about 0.5 to 0.95
Jupiter mass of gas, and ultimately become Neptune-mass planets.  Note that this evaporation
mechanism holds whether or not the planet has migrated inward from larger
orbits since its formation.  Migration implies the presence of a disk and upper
limits for disk lifetimes are $\sim 10$ Myr (Armitage et al. 2003).  Thus,
regardless of migration, the planet will spend the majority of its life in its
final short-period orbit, evolving under the influence of irradiation and
evaporation.

Given the large amount of mass lost by the planet, one may wonder about the consequences
on the planetary orbit. A tentative answer can be inferred 
from an analogy with the orbital evolution of comets.
If matter escapes isotropically, the planetary orbit remains unaffected. If
mass loss occurs anisotropically, however, the orbit can be affected. Like for comets, mass
loss will take place on the irradiated side of 
the planet and thus will occur initially in the direction of the parent star, exerting a force 
in the opposite direction and pushing the planet away from the star. 
This may affect the evaporation process and eventually quench it, if the orbital separation 
$a$ increases sufficiently. A quantitative estimate of this effect will depend on the fraction 
of anisotropy of the escaping flow and its velocity. Such a study is beyond the scope of the present paper. 
However, the existence of close-in planets on very short orbits ($a \, < \, 0.05$ AU)
suggests that this effect is small. This question definitly deserves a more thorough
analysis.

Note that in our model the upper atmosphere, where most of the incident XUV
flux is absorbed, always remains within the Roche lobe radius of the planet.
Above this limit, $\sim 1.3\,\rjup$ for $\rho^1$Cancri and $\mu$Ara, Roche lobe
overflow will occur, leading also to complete escape of the planet gas envelope
(Gu et al. 2003, Lecavelier et al. 2004).  The runaway of the evaporation process (Baraffe et al. 2004) leads to the catastrophic expansion of the planets, as indicated by the dotted lines in the bottom panel,
 and eventually to
the evaporation of the entire hydrogen/helium content.  The mass-radius relationships predicted by our
model are totally different from the one characteristic of a non-irradiated,
non-evaporating, H/He Neptune-mass planet which, to first approximation, would
have a radius $R\sim 0.6\,\rjup$ after 1 Gyr
(Zapolsky \& Salpeter 1969).  

\section{Discussion and conclusion}

Our evolutionary
models, accounting  for both irradiation and evaporation processes, show that most of the close-in exoplanets discovered today, including
the emerging Neptune-mass population, may originate from significantly more
massive Jupiter-like gas giants which have experienced drastic mass loss due to
the stellar high-energy radiation. 
This result is in direct opposition to the current belief
 that the hot-Neptunes are of different nature than the hot-Jupiters.  
This statement was used in a recent paper by Mazeh et al. (2004) who report
an interesting anti-correlation between mass and orbital period of transiting planets.
In their analysis, they ignored the recently discovered hot-Neptunes because of their supposedly different nature. We show here that this supposition is by no means obvious and that Jupiter-mass
and Neptune-mass close-in planets may share the same origin. Including the Neptune-mass
planets in the sample of all short period planets weakens the case for the anti-correlation of
Mazeh et al. (2004). 

Our scenario is certainly affected by the large uncertainties in the evaporation model 
of Lammer et al. (2003), which is based on the assumption that the planet undergoes {\it maximal}
energy-limited evaporation. This idea of energy-limited evaporation, meaning that the escape rate
is essentially determined by the amount of the high energy flux absorbed in the upper planetary atmosphere, was originally suggested by \"Opik (1963), and
further applied by 
Watson et al. (1981) and many others (see references in Lammer et al. 2003) 
to terrestrial planet atmospheres. As recently shown by Yelle (2004), energy-limited escape rates for
close-in giant planets may be limited by efficient cooling
due to  chemical species, such as H$_3^+$ and H$^+$, which decreases the efficiency of local heating due to the absorption of the XUV flux by the upper atmosphere. The absence of detailed chemistry in
the work by Lammer et al. (2003) may result in overestimated rates, as suggested by Yelle (2004).
Indeed, this latter finds escape rates 20 times smaller than the ones estimated by Lammer et al. (2003).
With such significantly smaller rates, hot-neptunes can not originate from Jupiter- or even Saturn-mass planets. Test calculations with an escape rate 10 times smaller than the one predicted by Lammer et al. (2003)
indicate that initial masses less than $\sim$ 0.1 $\mjup$ are required to reach a Neptune-mass 
with the orbital parameters of the three discovered hot-Neptune systems within a few Gyr . 
The formation of such low mass gaseous planets at such small
orbital distances seems to be very unlikely within the framework of current formation models of giant planets. This certainly would cast doubt on our present suggestion of a common origin
for close-in neptunian and jovian planets.

The work by Lammer et al. (2003) and Yelle (2004) are the first attempts to understand all the
complex processes involved in the evaporation of close-in giant planets. 
It is presently impossible to favor one of these models over the other. Since the Lammer et al. (2003) model
yields maximal energy-limited escape, it is important to examine its 
consequences on the planet fate as an upper limit case for the evaporation. The next step is to explore 
systematically the effect of lower evaporation rates, as well as of other uncertainties inherent
to our calculations, mainly the effect of initial conditions,  of a rocky/icy core and of
non-standard chemical composition in the envelope. This work is under progress.

The discovery of neptunian-mass {\it transiting} planets will be crucial for
testing the present scenario.  As shown in \S 3, 
 the proposed irradiation and evaporation global process will be
directly revealed by the planet mass-radius relationship.  As mentioned
previously, a purely gaseous (H/He) Neptune-mass planet not affected by
irradiation and evaporation during its evolution will have a radius $R\sim
0.6\,\rjup$ at present epoch ($>$ 1 Gyr).  Planets composed {\it dominantly} of
solid material, like terrestrial planets or the {\em ice} giants, Neptune and
Uranus, will have significantly more compact structures (Neptune's radius  is $\sim 0.35 \,\rjup$ at its present age).  In contrast, if irradiation plus
evaporation is a dominant process in the evolution of short-period exoplanets,
with Neptune-mass planets or even smaller objects originating from Jupiter-like
{\it gas giants}, we predict significantly larger radii for a given mass,
namely $R\gtrsim$ 0.8 $\rjup$ for the presently detected Neptune-mass objects,
as illustrated in Figure \ref{fig2}.  

The large radii we predict for Neptune mass planets place them well within the
detection limit of most transit surveys and the discovery of such an object
would strongly support our scenario. 
However, given the low probability of transit detections, no Neptune-mass transit discovery
remains consistent with the low number of transits detected up to now.
If no object with the presently predicted mass-radius relationships is found,
once larger transit statistics become available, then several possibilities
exist.  Perhaps the evaporation mechanism is not a dominant process in planet
history, questioning the observed evaporation of HD 209458b and/or the validity of the present
evaporation rates. Or, conceivably,
all the progenitors of Neptune-mass objects lie below the predicted critical
mass (see bottom panel of Figure \ref{fig2}) and, at most, a rocky core remains after
evaporation.  This latter case would suggest that many of the short-period
low-mass planets should be made mostly of rocks. 

On the other hand, transit detections of Neptune-mass
objects with Jupiter-like radii would superbly corroborate the present scenario
of irradiated and evaporating gaseous giants.  Our ability to find transiting planets
will improve significantly with the soon to be
launched COROT and KEPLER missions.  Comparisons between these observations and
the present calculations will potentially bring information about the
composition, evolution, and the origin of short-period exoplanets, unveiling
part of their mystery and possibly changing the understanding of our own Solar
system.


\begin{thebibliography}{}
\bibitem[]{} Armitage, P.J., Clarke, C. J., Palla, F. 2003, MNRAS, 342, 1139
\bibitem[]{} Baraffe, I., Chabrier G., Barman, T., Allard F., Hauschildt P
.H., 2003, A\&A, 402, 701
\bibitem[]{} Baraffe, I., Selsis, F., Chabrier G., Barman, T., Allard F., Hauschildt, P., 2004, A\&A, 419, L13
\bibitem[2001] {barman} Barman, T., Hauschildt, P.H., Allard, F. 2001, \apj, 
556, 885
\bibitem[] {} Barman, T.,  Allard, F.,  Hauschildt, P.H. 2005, in preparation
\bibitem[]{} Bouchy, F., Pont, F., Santos, N.C., Melo, C., Mayor, M., Queloz, D., Udry, S. 2004, A\&A, 421, L13
\bibitem[]{} Butler, R.P., Vogt, S.S., Marcy, G.W., et al. 2004, \apj, 617, 580
\bibitem[]{} Chabrier G., Barman, T., Baraffe, I.,  Allard F., Hauschildt, P., 2004, \apjl, 603, L53
\bibitem[]{} Charbonneau, D., Brown, T., Latham, D., Mayor, M. 2000,
\apj, 529, L45
\bibitem[2002]{cody} Cody, A.M., Sasselov, D.D. 2002, \apj, 569, 451
\bibitem[2002]{guillot} Guillot, T., Showman, A.P. 2002, \aap, 385, 156
\bibitem[]{} Gu, P., Lin, D.N.C., Bodenheimer, P.H. 2003, \apj, 588, 509
\bibitem[]{} Konacki, M., Torres, G., Jha, S., Sasselov, D., 2003, Nature, 421, 507
\bibitem[]{} Konacki, M., Torres, G.,  S., Sasselov, D., Jha, S. 2005, \apj, in press, (astro-ph/0412400)
\bibitem[]{} Lammer, H., Selsis, F., Ribas, I., Guinan, E., Bauer, S.J., Weiss, W.
2003, ApJL, 598, L121
\bibitem[]{} Lecavelier des Etangs, A., Vidal-Madjar, A., McConnell, J.C., H\'ebrard, G. 2004, A\&A, 418, L1
\bibitem[]{} Mazeh, T., Zucker, S., Pont, F. 2004, MNRAS, 356, 955
\bibitem[]{} McArthur, B.E., Endl, M., Cochran, W.D., et al. 2004, \apj, 614, L81
\bibitem[]{} Moutou, C., Pont, F., Bouchy, F., Mayor, M. 2004, A\&A, 424, L31
\bibitem[]{} \"Opik, E.J. 1963, Geophys. J. RAS, 7, 490 
\bibitem[]{} Pont, F., Bouchy, F., Queloz, D., Santos, N.C., Melo, C., Mayor, M., Udry, S. 2004, A\&A, 426, L15
\bibitem[]{} Ribas, I., Guinan, E.F., G\"udel, M., Audard, M. 2005, \apj, 622, 680
\bibitem[]{} Santos, N.C., Bouchy, F., Mayor, M., et al. 2004, A\&A, 426, L19
\bibitem[]{} Saumon D., Hubbard W.B., Burrows A., Guillot T., Lunine J.I.,
Chabrier, G., 1996, \apj, 460, 993
\bibitem[]{} Sozzetti, A., Yong, D., Torres, G., et al. 2004, \apj, 616, L167
\bibitem[]{} Torres, G., Konacki, M., Sasselov, D., Jha, S. 2004, \apj, 609, 1071
\bibitem[]{} Vidal-Madjar, A., Lecavalier des Etangs, A., D\'esert, J-M., Ballester, G.,
Ferlet, R., H\'ebrand, G., Mayor, M. 2003, Nature, 422, 143
\bibitem[]{} Vidal-Madjar, A., D\'esert, J-M., Lecavalier des Etangs, A., et al. 2004, \apj, 604, L69
\bibitem[]{} Watson, A.J., Donahue, T.M., Walker, J.C.G. 1981, Icarus, 48, 150
\bibitem[]{} Yelle, R.V. 2004, Icarus, 170, 167
\bibitem[]{} Zapolsky, H.S., Salpeter, E.E. 1969, \apj, 158, 809
\end{thebibliography}
\end{document}